\documentclass[aps,preprint,showpacs,preprintnumbers,amsmath,amssymb]{revtex4}

\usepackage{graphicx}
\usepackage{dcolumn}
\usepackage{bm}
\newcommand{\ybox}[2]   {
    \begin{center}
    \resizebox{!}{#1\textheight}
{\includegraphics{#2.eps}}
\end{center}           }
\def\refb#1{(\ref{#1})}
\begin{document}
\title{TeV black hole fragmentation and detectability in
extensive air showers}
\author{Eun-Joo Ahn}
    \altaffiliation{Email: sein@oddjob.uchicago.edu}
\affiliation{Department of Astronomy \& Astrophysics, University
of Chicago, 5640 S. Ellis Avenue, Chicago, IL 60637, USA}
\author{Maximo Ave}
     \altaffiliation{Email: ave@cfcp.uchicago.edu}
\affiliation{Enrico Fermi Institute, University of Chicago,
    5640 S.\ Ellis Avenue, Chicago, IL 60637, USA}
\author{Marco Cavagli\`a}
    \altaffiliation{Email: marco.cavaglia@port.ac.uk}
\affiliation{Institute of Cosmology and Gravitation, University of
Portsmouth, Portsmouth PO1 2EG, UK}
\author{Angela V.\ Olinto}
\altaffiliation{Email: olinto@oddjob.uchicago.edu}
\affiliation{Department of Astronomy \& Astrophysics, Enrico Fermi
Institute, and Center for Cosmological Physics, University
of Chicago, 5640 S.\ Ellis Avenue, Chicago, IL 60637, USA}
\date{\today}
\begin{abstract}
In models with large extra dimensions, particle collisions with center-of-mass
energy larger than the fundamental gravitational scale can generate
nonperturbative gravitational objects. Since cosmic rays have been observed
with energies above $10^{8}$ TeV, gravitational effects in the TeV energy range
can, in principle, be observed by ultrahigh energy cosmic ray detectors. We
consider the interaction of ultrahigh energy neutrinos in the atmosphere and
compare extensive air showers from TeV black hole formation and fragmentation
with standard model processes. Departures from the standard model predictions
arise in the interaction cross sections and in the multiplicity of secondary
particles.  Large theoretical uncertainties in the black hole cross section
weaken attempts to constrain TeV gravity based solely on differences between
predicted and observed event rates. The large multiplicity of secondaries in
black hole fragmentation enhances the detectability of TeV gravity effects. We
simulate TeV black hole air showers using PYTHIA and AIRES,  and find that
black hole-induced air showers are quite distinct from standard model
air showers. However, the limited amount of information registered by realistic
detectors together with large air shower fluctuations limit in practice the
ability to distinguish TeV gravity events from standard model events in a
shower by shower case. We discuss possible strategies to optimize the
detectability of black hole events and propose a few unique signatures that may
allow future high statistics detectors to separate black hole from standard
model events.
\end{abstract}
\maketitle
\section{Introduction}
In models with large extra dimensions, the fundamental scale of gravity may be
around TeV energies \cite{Antoniadis:1990ew, Arkani-Hamed:1998rs,
Antoniadis:1998ig, Randall:1999ee, Randall:1999vf}. The presence of extra
dimensions affects both sub- and super-Planckian physics. Sub-Planckian physics
is affected by the presence of Kaluza-Klein modes that lead to deviations from
standard model (SM) predictions in perturbative processes \cite{Giudice:1998ck,
Han:1998sg, Hewett:1998sn, Rizzo:1998fm}. Searches for these effects in
collider experiments have placed bounds on the fundamental Planck scale,
$M_{\star} \ge\, 1.3$ TeV for two extra dimensions and $M_{\star} \ge\, 0.25$
TeV for six extra dimensions \cite{Giudice:2003tu}. Additionally, submillimeter
tests of the gravitational inverse-square law constrain $M_{\star}\ge\, 1.6$
TeV for $n=2$ \cite{Adelberger:2002ic}.

Super-Planckian physics involves nonperturbative effects, the most striking
being the possible formation of black holes (BHs) \cite{Banks:1999gd} and other
gravitational objects \cite{Ahn:2002mj, Ahn:2002zn, Dimopoulos:2001qe,
Cheung:2002aq, Casadio:2001wh} in particle collisions with center-of-mass (CM)
energy larger than the fundamental Planck scale. (For recent reviews, see Refs.
\cite{Cavaglia:2002si, Tu:2002xs, Landsberg:2002sa, Emparan:2003xu}.) If the
fundamental scale is of the order of a few TeV, the products of BH decay could
be detected in particle colliders \cite{Giddings:2001bu, Dimopoulos:2001hw,
Cheung:2001ue, Mocioiu:2003gi} and in extensive air showers of ultrahigh
energies \cite{Feng:2001ib, Anchordoqui:2001cg, Anchordoqui:2001ei,
Ringwald:2002vk, Anchordoqui:2002hs, Dutta:2002ca}.

Ultrahigh energy cosmic rays provide a natural beam of particles with primary
energies up to and above $10^{8}$ TeV that can in principle probe TeV scale
physics. The dominant component of ultrahigh energy cosmic rays (UHECRs) is
believed to be protons \cite{Ave:2000nd} generated in extra-galactic sources.
UHECR protons naturally generate ultrahigh energy neutrinos as they traverse
intergalactic space through photo-pion production off the cosmic microwave
background (CMB) \cite{berezinsky1,berezinsky2}. The threshold energy for pion
production off the CMB induces a feature in the UHECR spectrum known as the
Greisen, Zatsepin, and Kuzmin (GZK) feature \cite{Greisen:1966jv,
Zatsepin:1966jv}. The flux of neutrino secondaries from the pion production
depends on the assumed extra-galactic proton injection spectrum  and generally
peaks around $10^6$ TeV \cite{berezinsky1, berezinsky2,Bhattacharjee:1998qc,
Engel:2001hd}. These secondary neutrinos are often called GZK or cosmogenic
neutrinos. Here we study the characteristics of extensive air showers initiated
by ultrahigh energy neutrinos and compare the production of BHs in TeV gravity
theories with SM interactions.

Ultrahigh energy neutrinos provide a useful means to test TeV gravity. In some
TeV gravity models, the neutrino-nucleon cross section, $\sigma_{\nu N}$, is
greatly enhanced, leading to larger numbers of neutrino-induced air shower
events. In fact, the lack of observed neutrino air showers can be used to place
a bound on the neutrino-nucleon cross section that has been translated into
constraints on $M_{\star}$ comparable to collider limits
\cite{Anchordoqui:2001ei, Tyler:2000gt}. However, the physics of BH formation
and evolution in TeV gravity theories is highly uncertain and model dependent.
As we discuss below, the cross section of the process can only be roughly
estimated. While some choices of parameters lead to the enhancement of
neutrino-nucleon cross sections compared to the SM, others choices give cross
sections for BH formation orders of magnitude below the SM case. Furthermore,
the evaporation process of BHs generates additional uncertainties on the
fraction of the primary energy that is left to generate a shower. Even if a
limit on the neutrino cross-section can be derived from the lack of
neutrino-induced air showers (for example, if the cosmogenic neutrino flux is
better constrained), translating a bound on $\sigma_{\nu N}$ into a limit on
TeV gravity parameters is highly model dependent. Therefore, the
identification of quantum gravity effects based solely on neutrino event rates
is not very effective.

Here we take a different approach by modeling the detailed
characteristics of extensive air showers initiated by BH evaporation on a
shower-to-shower basis, with the expectation that the large multiplicity of
secondaries will lead to detectable signatures. We first calculate the
fragmentation of BH and the spectrum of secondaries. The secondary particles
are then developed with PYTHIA \cite{Sjo01} and AIRES \cite{Sciutto:1999rr}
into observable extensive air showers. We find that BH-induced air showers
generally differ from ordinary air showers. Differences in shower maxima reach
$\sim 200$ g cm$^{-2}$  between BH and SM events which could be easily detected
if the first interaction point of the air showers were either observed or fixed
by the interaction. Unfortunately, the first interaction point of high energy
neutrinos in the atmosphere is neither fixed by the interaction nor detectable.
Unlike protons, the interaction length of neutrinos in air is quite large, thus
neutrinos interact with almost equal probability at any point in the
atmosphere.  Moreover, the first interaction point is not directly observed
since fluorescence experiments can only detect the air shower once billions of
particles have been generated while ground arrays only observe the air shower
as it reaches the ground. Shower observables such as the muon content and the
rise-depth parameter give indirect signatures that can distinguish BH and SM
events in large statistics experiments that combine fluorescence detectors and
ground array detectors.

In addition to differences in the overall characteristics of air showers, BH
formation produces some unique signatures since the fragmentation secondaries
span most particles in the SM. In particular, heavy BHs may produce several
$\tau$-leptons.  Multiple $\tau$'s are unique to BH formation and may be
differentiated in future UHECR observatories.

The paper is organized as follows. In the first part of \S II, we review BH
formation in TeV gravity and the physics of the BH-induced atmospheric events.
The aim of this part is to fix notations and make the paper self-consistent. In
the second part of \S II we first focus on the cross section uncertainties,
which make the identification of atmospheric BH formation based solely on event
rates ineffective. Next we discuss the phenomenology of BH evaporation, which
is the backbone of the air shower simulations. In \S III, we describe the Monte
Carlo that we have developed and used to simulate neutrino-induced air showers
in the atmosphere. The main results of the paper are contained in \S IV, where
we show the outcome of our simulations and discuss the differences between
ordinary air showers and BH-induced air showers. In \S V, we briefly discuss
possible detection techniques for BH formation based on $\tau$ production in BH
fragmentation. Finally, we conclude in \S VI.
\section{Black hole production in TeV gravity}
In models with $n$ extra dimensions the fundamental coupling constant of
gravity is the $(n+4)$-dimensional Newton's constant
\begin{equation}
G_{n+4}\,\equiv \, M_{\star}^{-(n+2)}\,.
\label{mstar}
\end{equation}
The observed four-dimensional Newton's constant $G_4 \equiv
M_{Pl}^{-2}=6.707 \times 10^{-33}$ TeV$^{-2}$ and the $(n+4)$-dimensional
gravitational constant $G_{n+4}$ are related by
\begin{equation}
G_4 \,=\, G_{n+4}V_{n}^{-1}\,,
\label{ggrel}
\end{equation}
where $V_{n}$ is the volume of the extra dimensions. If $V_{n}\gg
M_\star^{-n}$, it follows that $M_{\star}\ll M_{Pl}$. For the appropriate
choices of $n$ and $V_n$, $M_\star$ can be of the order of TeV energies such
that gravity and the electroweak scales coincide. These models provide an
attractive solution to the hierarchy problem of high-energy physics.

If gravity becomes strong at the electroweak scale, particle collisions with CM
energy larger than a TeV can create BHs \cite{Banks:1999gd}, branes
\cite{Ahn:2002mj,Ahn:2002zn}, and other nonperturbative gravitational objects
\cite{Dimopoulos:2001qe, Cheung:2002aq, Casadio:2001wh}. BH formation dominates
the gravitational channel if the extra-dimensional space is symmetric whereas
branes form in asymmetric cases \cite{Ahn:2002mj}. In this paper we only
consider symmetric compactification and BH production since brane decay is even
less understood than BH evaporation.
\subsection{Cross sections}
The static and uncharged BH in $(n+4)$-dimensions is described by the
$(n+4)$-dimensional Schwarzschild solution
\begin{equation}
ds^2 \,=\, -R(r)dt^2 \,+\, R(r)^{-1}dr^2 \,+\, r^2d\Omega ^2 _{n+2}\,,
\label{schwarz}
\end{equation}
where
\begin{equation}
R(r) \,=\, 1-\left(\frac{r_s}{r}\right)^{n+1}\,.
\label{R}
\end{equation}
The Schwarzschild radius $r_s$ of the BH is related to the mass $M_{BH}$ by
\begin{equation}
r_s=\frac{1}{\sqrt{\pi} M_\star}\left[\frac{8\Gamma\left(\frac{n+3}{2}\right)}
{(2+n)}\left(\frac{M_{BH}}{M_{\star}}\right)\right]^{\frac{1}{n+1}}\,.
\label{rmrel}
\end{equation}
At energy scales sufficiently above $M_\star$, BH formation is a semiclassical
process. Thus the cross section can be approximated by an absorptive black disk
with radius $r_s$. For a Schwarzschild BH the cross section is
\begin{equation}
\sigma_{ij \rightarrow BH}(s;n)\,=\,F(s)\pi r_s ^2 \,=\, F(s)\,
\frac{1}{s_{\star}} \left[\frac{8\Gamma\left(\frac{n+3}{2}\right)}
{(2+n)}\right]^{\frac{2}{n+1}}\left(\frac{s}{s_{\star}}
\right)^{\frac{1}{n+1}}\,,
\label{cross}
\end{equation}
where $\sqrt{s}$ is the CM energy of the collision, $s_{\star} =
M_{\star}^2$, and $F(s)$ is a form factor. Since $M_{BH}\gtrsim M_{\star}$,
it follows that $r_s \sim M_{\star}^{-1}$ and the cross section
\refb{cross} must be interpreted at the parton level. The total cross
section for a neutrino-proton event is obtained by summing over partons:
\begin{equation}
\sigma_{\nu p \to BH}(x_m;n) = \sum_{ij} \int_{x_m}^{1} dx \, q_i(x,-Q^2) \,
\sigma_{ij \rightarrow BH}(xs;n) \,,
\label{totcross}
\end{equation}
where $q_i(x,-Q^2)$ are the parton distribution functions (PDFs)
\cite{Brock:1993sz}, $-Q^2$ is the four-momentum transfer squared, $x$ is
the fraction of nucleon's momentum carried by the parton, and
$\sqrt{sx_m}=M_{BH,min}$ is the minimal BH mass for which the semiclassical
cross section is valid (generally $M_{BH,min}\sim$ few $M_\star$).

Equation \refb{totcross} should be interpreted with care as the total cross
section value is affected by several sources of uncertainty. The first
uncertainty comes from the approximate knowledge of the PDFs. For instance, the
uncertainty in the gluon distribution (the most uncertain distribution) is
$\sim$ 15\% for $x\lesssim 0.3$ and increases rapidly for large $x$
\cite{Pumplin:2002vw}. Furthermore, the PDFs are known only for momentum
transfer smaller than 10 TeV. In BH events we expect the momentum transfer to
be of the order of either the mass or the inverse Schwarzschild radius
\cite{Emparan:2001kf}. Therefore, the momentum transfer can reach hundreds of
TeV in UHECR-induced BH events. In the calculation of the total cross section
we fix the PDFs for momentum transfers above 10 TeV to be equal to the 10 TeV
value. Although the dependence of the PDFs on the momentum transfer seems quite
small (at least for momentum transfers smaller than 10 TeV), the 10 TeV cutoff
on the momentum transfer induces an additional uncertainty in the integrated
cross section Eq.\,\refb{totcross}. A conservative estimate of the total
uncertainty due to the PDFs is $\sim$ 20\%.

A second major source of uncertainty in Eq.\,\refb{totcross} derives from the
physics of BH formation at the parton level, which is presently not well
understood. The theoretical uncertainties in the dynamics of the process at
parton level are parametrized by the form factor $F(s)$ and have been
summarized in Refs.~\cite{Anchordoqui:2001cg,Cavaglia:2002si}. The two main
factors that may affect Eq.\,\refb{cross} are the uncertainty in the fraction
of the initial CM energy that goes into the BH and the presence of angular
momentum. Numerical simulations for head-on collisions in four dimensions
suggest that the mass of the BH is smaller than the CM energy of the colliding
particles, leading to a reduction of the total cross section. Rotating BHs have
also smaller cross sections than non-rotating BHs. A naive estimate of the
corrections due to angular momentum suggests a reduction of the cross section
of about 40\%. On the other hand, the non-relativistic limit of two-BH
scattering indicates that the geometrical cross section can be enhanced by a
factor $\sim 250-350$\%, depending on the spacetime dimension. The classical
cross section for photon capture can also be used to obtain a crude estimate of
the cross section of BH formation, suggesting an enhancement of the cross
section by a factor ranging from 300\% ($n=2$) to 87\% ($n=7$).

To our knowledge, all the quantitative results of the past literature are
obtained from Eq.\,\refb{totcross} by setting $F(s)=1$ and neglecting the PDF
uncertainties. (See, e.g., \cite{Feng:2001ib, Anchordoqui:2001cg,
Anchordoqui:2001ei, Ringwald:2002vk}.) This is partially motivated by the fact
that an exact estimate of the total uncertainty in the BH cross sections due to
the combined PDF and parton-level uncertainties is unattainable at present.
However, the arguments listed above suggest that the cross section
uncertainties range from $\sim$ 40\% to $\sim$ 300\% which can significantly
affect most results. Throughout this paper,  we take into account these
uncertainties when deriving observables.

\begin{figure}
\ybox{0.7}{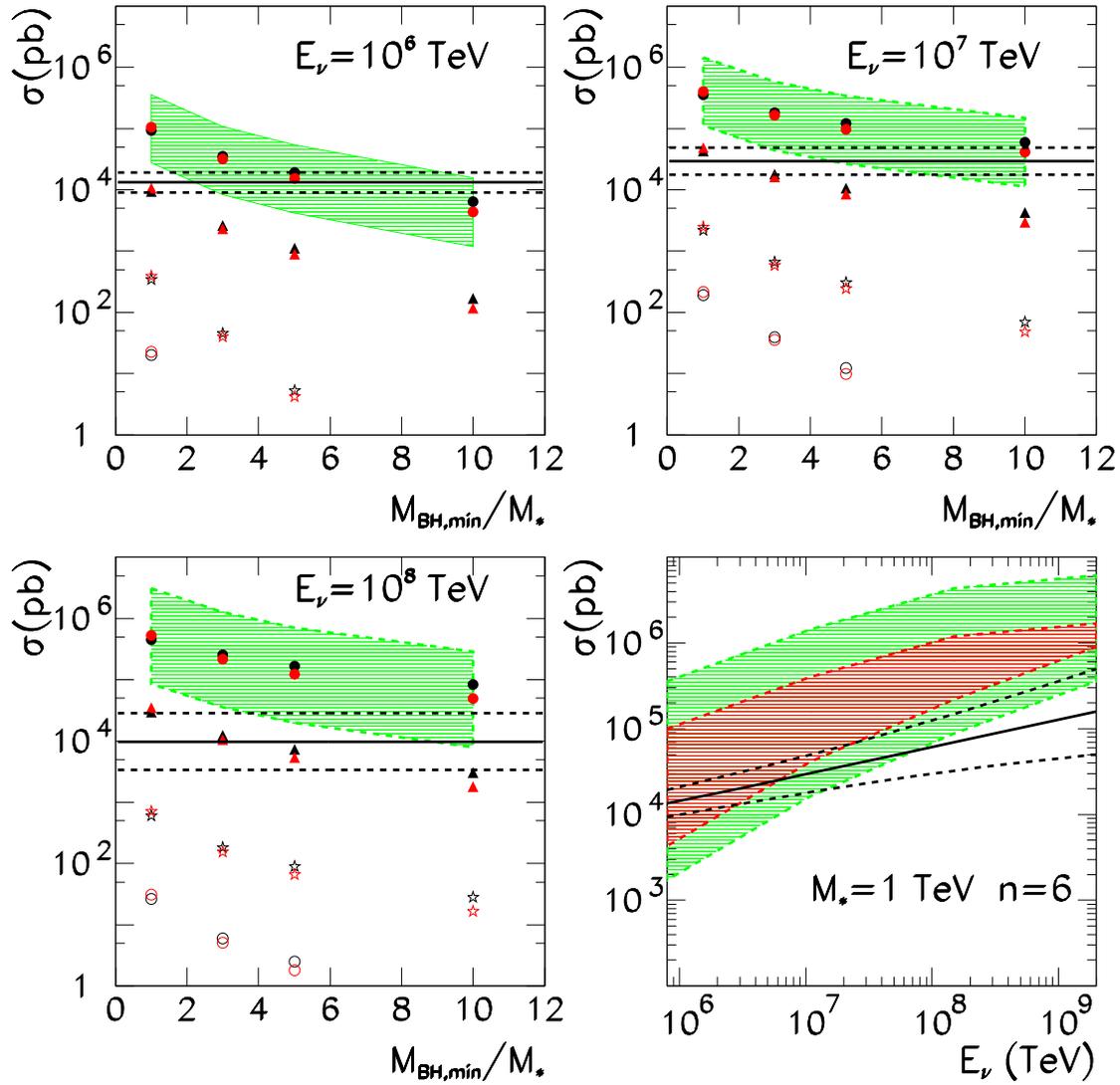}
\caption{The two upper panels and the lower left panel show $\sigma_{\nu p \to
BH}(M_{BH,min}/ M_\star)$ for $E_{\nu} = 10^6, 10^7, and 10^8$ TeV. The disks,
triangles, stars, and circles are for $M_{\star}=1$, $2$, $5$, and $10$ TeV,
respectively. The red (black) symbols are for $n=3$ ($n=6$). The shaded regions
show the uncertainties. The lower right panel shows $\sigma_{\nu p \to
BH}(E_{\nu})$ for $n=6$ and $M_{\star}=1$ TeV, with the $M_{BH,min}$ range in
red (inner shaded region) and the uncertainties at the parton level and PDF in
green (outer shaded region). The solid lines give the SM cross section, with
dashed lines showing PDF uncertainties.}
\label{crossfigure1}
\end{figure}

The cross section as a function of $M_\star$, $M_{BH,min}$, $E_\nu$, and the
uncertainties described above are shown in Fig.\,\ref{crossfigure1}. The upper
left, upper right, and lower left panels show $\sigma_{\nu p \to BH}$ as a
function of $M_{BH,min}$ for three different incoming neutrino energies ($E_\nu
= 10^6,$ $10^7$, and $10^8$ TeV), for different values of $M_\star$ (disks for
$M_\star =$ 1 TeV, triangles for $M_\star =$ 2 TeV, stars for $M_\star =$ 5
TeV, and circles for $M_\star =$ 10 TeV), and for the case of seven dimensions
($n=3$, red symbols) and ten dimensions ($n=6$, black symbols). The symbols
give the cross section calculated from Eq.\,\refb{totcross} setting $F(s)=1$
and neglecting the PDF uncertainties. The lower right panel shows the cross
section as a function of energy for $M_\star=1$ TeV and ten dimensions. The red
shaded (inner) region shows the uncertainty in the cross section due to the
unknown $M_{BH,min}$ which we vary from $M_\star$ to 10 $M_\star$. The green
shaded (outer) region shows the uncertainty associated with the BH formation at
the parton level and the PDF. The solid lines in the graphs show the cross
section for SM interactions. The uncertainty in the SM cross section due to the
unknown PDFs at very small values of $x$ is bracketed by dashed lines.

For a given number of extra dimensions, the total cross section increases with
the energy of the primary neutrino and decreases when the fundamental scale is
increased. At fixed energy and $M_\star$, the cross section decreases with
increasing $M_{BH,min}$. The overall effect makes the cross section at fixed
energy vary by many orders of magnitude. For instance, the ten-dimensional
total cross section at $E_\nu=10^8$ TeV spans five orders of magnitude, ranging
from values of tens of pb to millions of pb, where the lower values are
obtained for large fundamental scales.  At fixed energy and $M_\star$, the
range in cross section values span about an order of magnitude unless
$M_{BH,min}$ becomes comparable to the CM energy of the event. In this case,
the rate of events is dramatically suppressed and the cross section tends to
zero. The behavior of the total cross section with the energy at fixed
$M_\star$  is steeper for higher values of $M_{BH,min}$. In the range
$E_{\nu}=10^6-10^9$ TeV, which is of interest to UHECRs, the cross section
grows approximately like $\sigma_{\nu p \to BH}\sim E_\nu^{0.4\,-\,1.8}$, where
lower exponents are obtained for lower $M_{BH,min}$. For example, in ten
dimensions the behavior of the cross section is $\sigma_{\nu p \to BH}\sim
E_\nu^{0.41}$ for $M_{BH,min}=M_\star=1$ TeV, $\sigma_{\nu p \to BH}\sim
E_\nu^{0.67}$ for $M_{BH,min}=5\,M_\star=5$ TeV, and $\sigma_{\nu p \to BH}\sim
E_\nu^{0.77}$ for $M_{BH,min}=10\,M_\star=10$ TeV.

The large uncertainties in the values of  $\sigma_{\nu p \to BH}$ make it quite
difficult to discriminate between different values of the fundamental scale
$M_{\star}$ with good precision. The range of possible $\sigma_{\nu p \to BH}$
for a given  $M_{\star}$ overlaps with the range for larger $M_{\star}$ because
of the theoretical uncertainties. Even if $\sigma_{\nu p \to BH}$ were to be
constrained by experiments, $M_\star$ could not be determined unless the
degeneracy were removed by additional assumptions on $M_{BH,min}$ and by
reducing  other theoretical uncertainties in $\sigma_{\nu p \to BH}$. The
dependence of the cross section on $n$ is the least dramatic making it also
hard to differentiate between different dimensions.  In addition, as
$M_{\star}$ becomes larger than $\sim 1$ TeV, $\sigma_{\nu p \to BH}$ becomes
smaller than the cross section for SM interactions and the probability for BH
formation decreases accordingly.
\subsection{BH evaporation products}
\begin{figure}
\ybox{0.35}{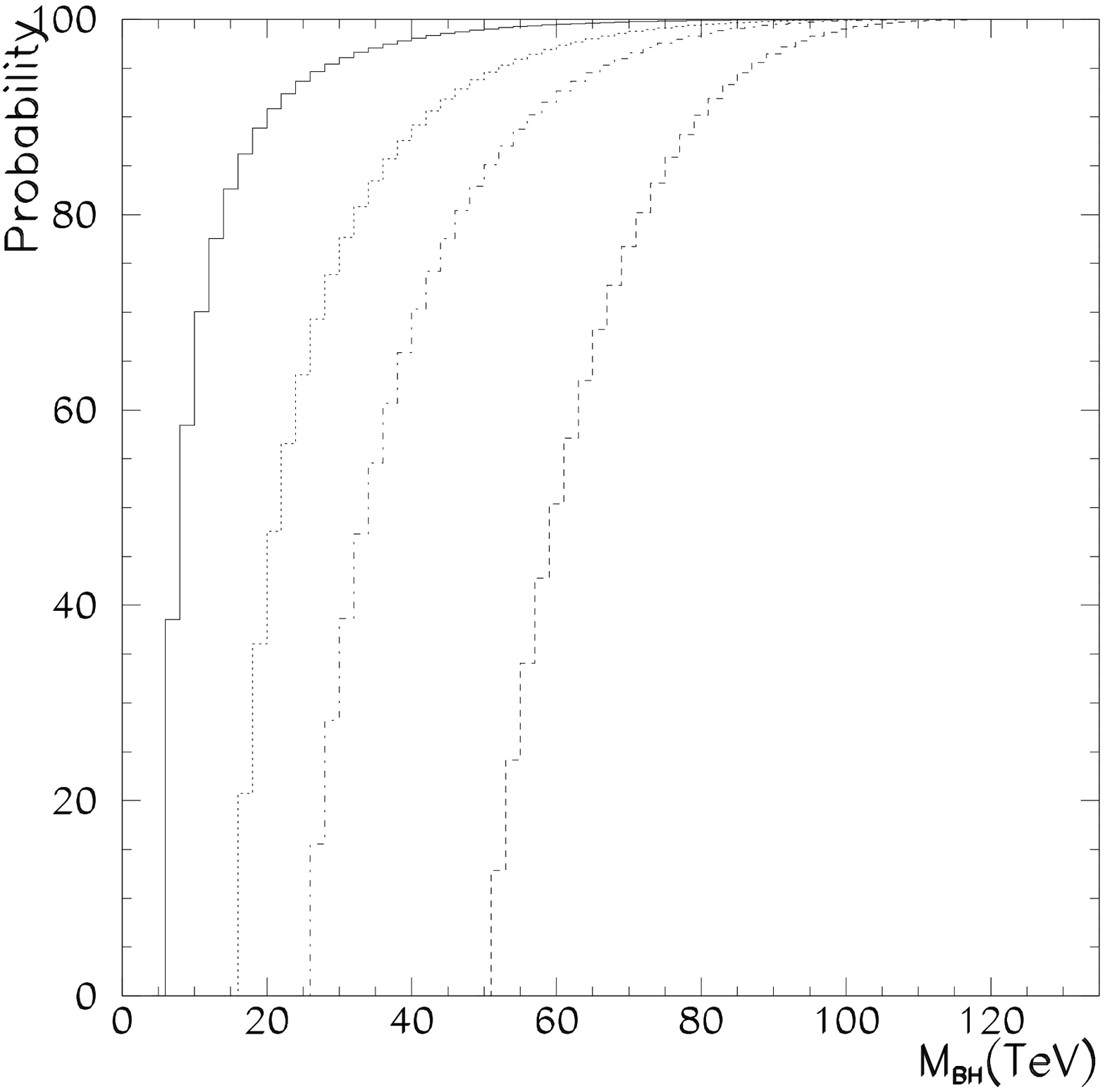}
\caption{Integrated probability of BH formation as a function of the initial BH
mass for $E_{\nu} = 10^7$ TeV, $n=6$, $M_\star=5$ TeV, and $M_{BH,min}=$ 1
(solid line), 3 (dotted line), 5 (dashed line), and 10 (long-dashed line)
$M_\star$.}
\label{probmass}
\end{figure}

Once the CM energy of a neutrino collision with a nucleon in the air reaches
the  BH formation threshold, a BH with initial mass equal to a fraction of the
total CM energy may form. The distribution of the initial BH masses is given by
the differential cross section
\begin{equation}
\frac{d\sigma_{\nu p \to BH}}{dM_{BH}} = 2\left(\frac{x}{s}\right)^{1/2}
q_i(x,-Q^2) \, \sigma_{ij \rightarrow BH}(xs;n) \,.
\label{diffcross}
\end{equation}
Light BHs are favored over heavy BHs. The typical initial BH mass is usually a
few times $M_{BH,min}$. Therefore, models with larger (smaller) fundamental
Planck scale tend to produce higher (lower) mass BHs. The integrated
probability of BH formation vs the initial BH mass is plotted in
Fig.\,\ref{probmass} for a neutrino energy $E_\nu=10^{7}$ TeV, $n=6$,
$M_\star=5$ TeV and $M_{BH,min}=1,3,5,10\,M_\star$. The initial BH mass is very
sensitive to the value of $M_{BH,min}$. For $M_{BH,min}=M_\star$, 90\% of the
formed BHs have initial mass less $\sim 20$ TeV, whereas for
$M_{BH,min}=10\,M_\star$ the 90\% threshold is reached at $M_{BH}\sim 80$ TeV.

Once formed, the BH decay phase is expected to proceed in three stages:
classical, semiclassical, and quantum \cite{Giddings:2001bu}. In the first
stage the BH sheds the hair associated with gauge charges and angular momentum.
In the second stage the BH evaporates semiclassically by emission of thermal
Hawking radiation with temperature $T_H$. We assume that most of the energy is
radiated into the brane, as only gravitons can ``see'' the bulk
\cite{Emparan:2000rs, Cavaglia:2003hg}. The Hawking evaporation ends when the
mass of the BH approaches $\sim M_\star$. At this point the semiclassical
description breaks down and the BH may either decay completely by emitting a
few quanta with energy of order of $M_\star$ or leave a stable remnant with
mass $\sim M_\star$ \cite{Cavaglia:2003qk}. The details of this last stage
depends on the unknown underlying quantum theory. However, the semiclassical
decay should lead to most of the observable signatures.
\begin{table}
\caption{Decay time in the CM frame ($\tau$, in units of $10^{-26}$ \,s),
initial temperature ($T_H$ in TeV), initial entropy ($S$), average number of
produced quanta ($\langle N \rangle$), and energy per quantum ($E/\langle N
\rangle$ in TeV) of Schwarzschild BHs for various $M_{\star}$ (TeV),
$M_{BH}$ (TeV), $n=6$ and $n=3$ (in parentheses).}
\begin{center}
\begin{tabular}{|c|c||ccccc|}
\hline
\small $M_{\star}$  & \small $M_{BH}$ & $\tau$  &
$T_H$  & $S$ & $\langle N \rangle $ & $E/ \langle N \rangle$ \\
\hline
\small 1 & \small 5 & \small 0.521 (0.736)& \small 0.553 (0.282) & \small 8
(14) & \small 5 (9) & \small 1.11 (0.56)\\
\small 1 & \small 10 & \small 1.27 (2.08)& \small 0.500 (0.237)& \small
17 (34) & \small 10 (21)  & \small 1.00 (0.47) \\
\small 1 & \small 50 & \small 10.1 (23.3) & \small 0.398 (0.159)& \small
110 (252) & \small 63 (158) & \small 0.80 (0.32) \\
\small 1 & \small 100 & \small 24.5 (65.8) & \small 0.360 (0.133)& \small
243 (600) & \small 139 (375) & \small 0.72 (0.27) \\

\hline
\small 2 & \small 5 & \small 0.107 (0.130) & \small 1.22 (0.671)
& \small 4 (6) & \small 2 (4) & \small 2.44 (1.34)\\
\small 2 & \small 10 & \small 0.261 (0.368) & \small 1.11 (0.564) & \small
8 (14) & \small 5 (9) & \small 2.21 (1.13) \\
\small 2 & \small 50 & \small 2.06 (4.11) & \small 0.878 (0.377) & \small
50 (106) & \small 28 (66) & \small 1.76 (0.76) \\
\small 2 & \small 100 & \small 5.03 (11.6) & \small 0.795 (0.317) & \small
110 (252) & \small 63 (158) & \small 1.59 (0.64) \\
\hline

\small 5 & \small 5 & \small 0.013 (0.013) & \small 3.48 (2.11) & \small
1 (2) & \small 1 (1) & \small 6.95 (4.22) \\
\small 5 & \small 10 & \small 0.032 (0.037) & \small 3.15 (1.77) & \small
3 (5) & \small 2 (3) & \small 6.30 (3.55)\\
\small 5 & \small 50 & \small 0.254 (0.416) & \small 2.50 (1.19) & \small
17 (34) & \small 10 (21) & \small 5.01 (2.37)\\
\small 5 & \small 100 & \small 0.620 (1.18) & \small 2.27 (0.997) & \small
39 (80) & \small 22 (50) & \small 4.53 (2.00) \\
\hline

\small 10 & \small 10 & \small 0.007 (0.007)& \small 6.95 (4.22)& \small
1 (2) & \small 1 (1)  & \small 13.9 (8.44) \\
\small 10 & \small 50 & \small 0.052 (0.074) & \small 5.53 (2.82)& \small
8 (14) & \small 5 (9) & \small 11.1 (5.64) \\
\small 10 & \small 100 & \small 0.127 (0.208) & \small 5.01 (2.37)& \small
17 (34) & \small 10 (21) & \small 10.0 (4.74) \\

\hline
\end{tabular}
\end{center}
\label{numbertable1}
\end{table}
During the semiclassical evaporation, the BH decays in a time (CM frame)
\cite{Cavaglia:2003hg}
\begin{equation}
\tau
\sim\frac{1}{M_{\star}}\left(\frac{M_{BH}}{M_{\star}}\right)^{\frac{n+3}{n+1}}
\,.
\label{lifetime}
\end{equation}
Assuming a Boltzmann statistics and instantaneous BH evaporation, the BH emits
an average number of quanta \cite{Dimopoulos:2001hw}
\begin{equation}
\langle N \rangle =\frac{M_{BH}}{2\,T_H}\,,
\label{timenum}
\end{equation}
where the Hawking temperature $T_H$ is related to the Schwarzschild radius
and to the entropy of the BH, $S_{BH}$, by \cite{Argyres:1998qn}
\begin{equation}
T_H = \frac{n+1}{4\pi r_s}=\frac{n+1}{n+2}\frac{M_{BH}}{S_{BH}}\,.
\label{tempent}
\end{equation}
In Table\,\ref{numbertable1}, we list the parameters of typical BHs in ten and
six dimensions for different choices of $M_{\star}$ and of the BH mass
($M_{BH}=$5, 10, 50, and $100$ TeV). The particle emission rate for a BH with
temperature $T_H$ is given by \cite{Han:2002yy, Anchordoqui:2002cp}
\begin{equation}
\frac{dN_i}{dE dt} \,=\, \frac{c_i \,\Gamma_{s_i}A_c}{8 \pi^2}\frac{E^2}
{e^{E/T}-(-1)^{2s_i}}\,,
\label{pageemi}
\end{equation}
where $E$ is the energy, $A_c$ is the optical area of the BH
\cite{Emparan:2000rs}, and $\Gamma_i$, $c_i$, $\sigma_{s_i}$, and $N_i$ are the
spin, the degrees of freedom, the greybody factors, and the number of quanta of
the particle of species $i$. We neglect particle masses which are generally
much smaller than the BH mass. Integrating Eq.\,\refb{pageemi} over $E$ gives
\begin{equation}
\frac{dN_i}{dt}=c_i\Gamma_{s_i}f_i\frac{AT^3}{8\pi^2}
\Gamma(3) \zeta(3)=c_i\Gamma_{s_i}f_i\, \frac{\zeta(3)T}{16
\pi^3}\frac{(n+3)^{\frac{n+3}{n+1}} (n+1)}{2^{\frac{2}{n+1}}}\,,
\label{hanemi}
\end{equation}
where $f_i=1$ ($3/4$) for bosons (fermions). Since the observed Hawking
emission happens on the brane, we use the four-dimensional greybody factors of
Ref.\,\cite{Page:df}. (Greybody factors in $n+4$ dimensions have been
calculated in Refs.\,\cite{Kanti:2002nr, Kanti:2002ge, Park:2002ez}. See also
Ref.\, \cite{Frolov:2002as, Frolov:2002gf} for a discussion on BH recoil
effect.) The values of $s_i$, $c_i$, $\Gamma_{s_i}$, and $f_i$ are listed in
Table\,\ref{emivalue}.

The number ratio of two particle species $i$ and $j$ is \cite{Cavaglia:2003hg}:
\begin{equation}
\frac{N_i}{N_j}=\frac{c_i\Gamma_{s_i}f_i}{c_j\Gamma_{s_j}f_j} \,.
\label{ratioeqn1}
\end{equation}
Using Eq.\,\refb{timenum}, $N_i$ can be expressed as
\begin{equation}
\displaystyle
N_i=\langle N \rangle\frac{c_i\Gamma_{s_i}f_i}{\sum_{j} c_j\Gamma_{s_j}f_j
}\,.
\label{ratioeqn4}
\end{equation}
The number of each particle species formed for the BHs listed in Table
\ref{numbertable1} for $M_{\star}=1$ TeV is given in Table\,\ref{caviatable1}.

\begin{table}[h]
\caption{The values of $s_i$, $c_i$,  $\Gamma_{s_i}$, $f_i$ for the SM
particles}
\begin{center}
\begin{tabular}{|c||c|c|c|c|}
\hline
\small species & \small $s_i$ & \small $c_i$ & \small $\Gamma_{si}$ & \small
$f_i$ \\
\hline
\small quark & \small 1/2 & \small 72 & \small 0.6685 & \small 3/4 \\
\small charged lepton & \small 1/2 & \small 12 & \small 0.6685 &\small 3/4\\
\small neutrino & \small 1/2 & \small 6 & \small 0.6685 & \small 3/4 \\
\small Higgs & \small 0 & \small 1 & \small 1 & \small 1  \\
\small photon & \small 1 & \small 2 & \small 0.2404 & \small 1 \\
\small gluon & \small 1 & \small 24 & \small 0.2404 & \small 1  \\
\small $W$ & \small 1 & \small 6 & \small 0.2404 & \small 1  \\
\small $Z$ & \small 1 & \small 3 & \small 0.2404 & \small 1  \\
\small graviton & \small 2 & \small 2 & \small 0.0275 & \small 1 \\ \hline
\end{tabular}
\end{center}
\label{emivalue}
\end{table}

\begin{table}[h]
\caption{Fragmented number of particle species for $n=6$ ($n=3$), $M_{\star} =
1$ TeV}
\begin{center}
\begin{tabular}{|c||c|c|c|c|}
\hline
\small $M_{BH}$ (TeV) & \small 5 & \small 10 & \small 50 & \small 100 \\
\hline
\small quark & \small 3 (6) & \small 7 (14) & \small 42 (104) & \small
92 (248)\\
\small c. lepton & \small 0 (1) & \small 1 (2) & \small 7 (17) & \small
15 (41)\\
\small neutrino & \small 0 (0) & \small 1 (1) & \small 3 (9) & \small 8
(21) \\
\small Higgs & \small 0 (0) & \small 0 (0)  & \small 1 (3) & \small 3
(7)\\
\small photon & \small 0 (0) & \small 0 (0) & \small 1 (1) & \small 1 (3)
\\
\small gluon & \small 0 (1) & \small 1 (2) & \small 7 (17) & \small 15
(40) \\
\small $W$ & \small 0 (0) & \small 0 (1) & \small 2 (4) & \small 4 (10) \\
\small $Z$ & \small 0 (0) & \small 0 (0) & \small 1 (2) & \small 2 (5) \\
\small graviton & \small 0 (0) & \small 0 (0) & \small 0 (0) & \small 0 (0)
\\
\hline
\end{tabular}
\end{center}
\label{caviatable1}
\end{table}

For example, a BH of mass $M_{BH} = 50$ TeV and $M_{\star} = 1$ TeV according
to Table\,\ref{numbertable1} emits 63 quanta, each with energy of 0.80 TeV.
These quanta are translated into SM particles some of which decay or hadronize.
The final output of the BH evaporation may contain up to $\sim$ 2000 particles.

\section{Extensive air shower simulations}
Extensive air showers created by ultrahigh energy interactions in the
atmosphere can be detected with ground arrays and fluorescence telescopes.
Ground arrays record the signal which is produced by the particles of the
shower reaching the ground. Fluorescence telescopes observe the fluorescence
light produced by the interaction of the atmospheric nitrogen molecules with
the electromagnetic component of the developing air shower. The fluorescence method
pioneered by the Fly's Eye \cite{Bird:yi,Bird:1994uy} detector is able to
reconstruct the longitudinal development of the $e^+e^-$ component of the
air shower. Fluorescence detectors are currently used by the HiRes
\cite{Abu-Zayyad:uu} and Auger \cite{Auger} experiments and are planned for the
future EUSO \cite{Euso} and OWL \cite{owl} observatories. This technique
provides a good estimate of the energy of the primary particle that initiate
the air shower, since most of the energy of the air shower goes into the observable
electromagnetic channel. Another advantage of the fluorescence technique is the
ability to reconstruct the depth at which the cascade contains the maximum
number of $e^+e^-$ pairs, i.e., the depth of shower maximum, $X_m$. This
parameter is sensitive to the type of primary particle, to its energy and to
the interaction initiating the cascade. The depth of the first interaction
point, $X_0$, depends on the total cross section of the particle considered.
Due to the small values of the neutrino-air cross sections, ultrahigh energy
neutrinos can induce air showers at any depth in the atmosphere such that $X_0$ is
arbitrary.

\begin{figure}
\ybox{0.35}{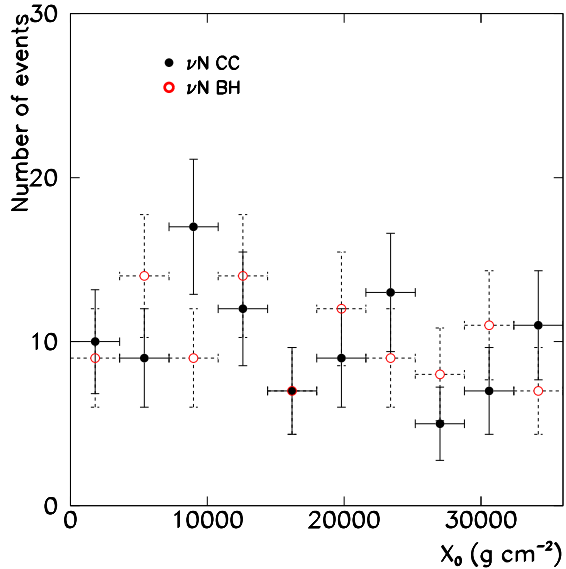}
\caption{The $X_0$ distribution for 100 neutrinos with  $E_\nu=10^{9}$ TeV
interacting in a column depth of $3.6 \times 10^4$ g~cm$^{-2}$. The SM CC
interaction length is $\lambda_{\nu CC} \simeq  1.1 \times 10^7$
g~cm$^{-2}$ (solid error bars). The BH interaction length is
$\lambda_{\nu BH}
\simeq 1.7\times10^5$ g~cm$^{-2}$ for $M_{\star}=1$ TeV and $n=6$ (dashed
error bars).}
\label{xmax}
\end{figure}

Neutrinos can interact at any depth in the atmosphere with almost equal
probability. The interaction length of a neutrino with energy $E_\nu=10^{9}$
TeV is $\lambda_{\nu CC} \simeq 1.1 \times 10^7$ g~cm$^{-2}$ for the charged
current (CC) interactions. This is larger than the column depth of the
atmosphere in the horizontal direction,  which is $3.6 \times 10^4$
g~cm$^{-2}$. BH forming interactions do not improve this situation as the BH
formation cross sections cannot be not much greater than the SM values (see
Fig.\,\ref{crossfigure1}). For example, if  $M_{\star}= 1$ TeV and $n=6$,
$\lambda_{\nu BH} \simeq 1.7\times10^5$ g~cm$^{-2}$. Fig.\,\ref{xmax} shows
that the $X_0$ distribution is flat for SM and BH interactions. Thus the $X_m$
distribution is also flat. As we discuss below, differences between SM and BH
interactions are evident in $X_m-X_0$. We can directly compare the values of
$X_m$ by fixing the value of $X_0$ in the simulations.

We developed a Monte Carlo code to study the air showers induced by BH formation in
neutrino-air collisions and compare the BH-induced air showers to the SM
neutrino-induced air showers. The code generates observable secondaries from SM
neutrino interactions and BH evaporation using the PYTHIA (ver.~6.2) package
\cite{Sjo01}. These secondaries are then injected into the AIRES simulator as
primaries for the final air shower. In the AIRES code the threshold energy for
tracking particles in the air showers are 80 keV for gamma rays, 80 keV for
electrons and positrons, 1 MeV for muons, 500 keV for mesons and 150 MeV for
nucleons.  The geomagnetic field is set to the Pierre Auger Observatory. The
``thinning'' level used in this work is 10$^{-5}$ with a weight limitation of
20. (Thinning is a method commonly used in simulations of UHECRs to avoid
following the huge number of secondary particles by following only a fraction
of them with varying weights. See \cite{Sciutto:1999rr} for further details.)
\subsection{SM neutrino-induced air showers}
We simulate the air showers induced by CC and neutral current (NC)
interactions by the following procedure:
\begin{itemize}
\item The differential cross sections are integrated over the fraction of
the total momentum of the nucleon carried by the parton ($x$) for all the
possible values of the fraction of total energy that goes into the hadronic
cascade ($y$).
\item A value of $y$ is sampled from the previous distribution. The mean
value of $y$ at the energies relevant for UHECRs is $0.2$.
\item The energy of the lepton (CC interaction) or neutrino (NC interaction) in
the final state is given by $(1-y)\,E_\nu$. The CC lepton is injected into AIRES.
$\tau$ leptons cannot be simulated by AIRES. Therefore, we calculate the decay
length and use the PYTHIA generator to obtain the secondaries, which are then
injected at the corresponding height at which the $\tau$ decays. Note, however,
that the $\tau$ particles have a mean energy of 0.2$E_\nu$, so most $\tau$'s
reach the ground without decaying and are not converted into observable energy.
The NC neutrino is not observable and is not injected in AIRES.
\item The hadronic part of the CC and NC interactions are simulated with
PYTHIA. The secondary particles are then injected in AIRES.
\end{itemize}
\subsection{BH-induced air showers}
A similar Monte Carlo code is used to simulate air showers induced by BH
formation. The BH simulation follows this procedure:
\begin{itemize}
\item The mass of BH is calculated using the probabilities given by
Eq.\,\refb{diffcross}. The gamma factor of the BH is $\gamma=E_\nu$/$M_{BH}$.
Different cases of $M_{BH,min}$ are considered.
\item The temperature of the BH, and the energy and total number of quanta
emitted in the evaporation phase are calculated for different choices of $n$
and $M_\star$. The fragmented number of particles species is computed (as in
Table\,\ref{caviatable1}).
\item The momentum of each quanta in the rest frame of the BH are calculated
assuming an isotropic distribution. If the quantum generated is a quark or a
gluon, the secondaries resulting from the parton cascade of this quantum are
calculated using PYTHIA. If the quantum is a gauge boson, it is decayed
using PYTHIA. The momenta of all the particles are then boosted to the
laboratory system.
\item The secondaries are injected in the AIRES code to simulate the
extensive air shower. All the secondaries are injected at the assumed first
interaction point except for the $\tau$ particles which are dealt with as in
the SM air showers. However, the energy of the $\tau$ particles generated by BH
evaporation is generally smaller than the energy of the $\tau$'s generated
in the SM process. The decay length of BH $\tau$'s is comparatively shorter
than in the SM case. Neutrinos, gravitons, and $\tau$'s that decay after
reaching the ground are not observable, thus they and are not injected in
AIRES.
\item $\pi^0$'s generated by the hadronization of quarks, gluons, and gauge
bosons are immediately decayed by PYTHIA. This is a good approximation since
the average pion energy is smaller than the critical energy. Therefore,
$\pi^0$'s are more likely to decay than interact.
\end{itemize}
\section{Simulation results}
\subsection{Neutrino-initiated air showers}
We simulated SM-induced air showers for CC and NC interactions as well as
air showers from BH production. The showers were chosen to have a zenith angle
of $70^\circ$ and a primary neutrino energy $E_\nu=10^{7}$ TeV. The first
interaction point was fixed to an altitude of 10 km corresponding to a slant
depth of 780 g cm$^{-2}$.

SM neutrino air showers are generally dominated by CC interactions because NC
interactions have lower cross section, $\sigma_\nu^{NC}=\,0.4
\,\sigma_\nu^{CC}$. Moreover, a large fraction of the primary neutrino energy
of the NC interaction, $(1-y)\,E_\nu \sim 0.8\, E_\nu $, is carried out by the
scattered neutrino and is not observable. Similarly, the CC $\nu_\mu$
scattering produces a high energy invisible $\mu$ that does not contribute to
the shower energy. The CC $\nu_\tau$ interaction produces a high energy $\tau$
that generally does not decay before reaching ground level. For $E_\nu=10^7$
TeV, the decay length of the scattered $\tau$ is $\sim$ 500 km. If the $\tau$'s
were to decay before reaching the ground, the air shower would appear as a
superposition of showers initiated at different heights. (We will return to
$\tau$ decay later in \S V.) Therefore, as far as CC interactions are
concerned, the most easily observed primary is the $\nu_e$. The secondary
electron initiates a purely electromagnetic air shower that carries $\sim$ 80\% of
the primary neutrino energy. These showers have similar features to
electromagnetic air showers:
\begin{itemize}
\item CC $\nu_e$ air showers are $\mu$-poor. The dominant process for $\mu$
production in an electromagnetic cascade is photoproduction. The ratio of the
pair production and  photoproduction cross sections determines the number of
$\mu$'s in the air shower. This ratio is $2.8 \times 10^{-3}$ at $10^{-2}$ GeV and
is expected to be $\sim 10^{-2}$ at $\sim 10^{7}~$ TeV.
\item CC $\nu_e$ air showers develop slower than hadronic air showers. The number of
secondaries in pair production or bremsstrahlung interactions is smaller than
in hadronic interactions. Additionally, the Landau-Pomeranchuk-Migdal (LPM)
effect \cite{Landau:um,Landau:gr,Migdal:1956tc,Migdal:1957} also contributes in
slowing down the shower development once primary energies reach above
$E_{LPM}\sim 10^{7}$ TeV \cite{aharonian,Cillis:1998hf}.
\end{itemize}

\begin{figure}
\ybox{0.35}{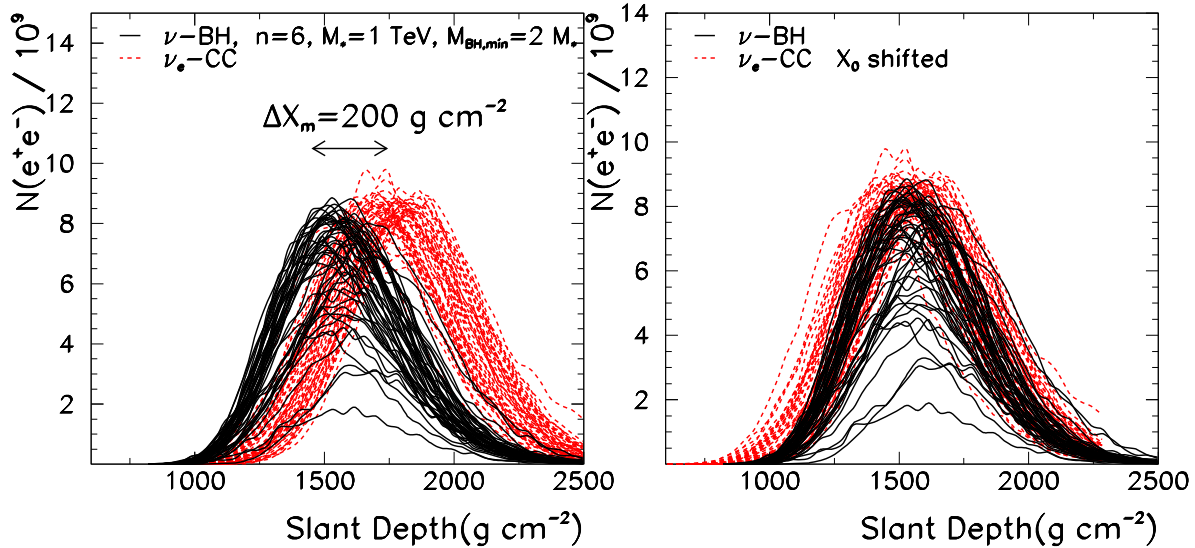}
\caption{Number of $e^+e^-$ vs slant depth for neutrino air showers with
$E_\nu=10^7$ TeV. The SM CC air showers are shown in red (dashed lines) and the
BH-induced air showers with $n=6$ and $M_{BH,min}=2M_{\star}=2$ TeV are shown
in black (solid lines). The left panel has fixed $X_0(CC)=X_0(BH)$. The right
panel has shifted $X_0$ such that $X_m(CC) \simeq X_m(BH)$.}
\label{longi1}
\end{figure}

We performed a systematic simulation of SM neutrino-induced air showers and
checked the characteristics discussed above. Here, we only show the more
relevant CC-induced air showers and compare them to the BH-induced air showers.
The left panel in Fig.\,\ref{longi1} shows the longitudinal development of CC-
and BH-induced air showers. We chose the BH parameters
$M_{BH,min}=2\,M_{\star}=2$ TeV and $n=6$. A difference of $\sim 200$ g
cm$^{-2}$ in $X_m$ is evident between the SM and BH events. This large
difference results from the combination of the large multiplicity and hadronic
nature of BH-induced air showers and electromagnetic nature of the CC-induced
air showers.

BH-induced air showers generally develop faster than typical SM hadronic
air showers depending on the initial and minimum masses of the BH that give the
number of produced quanta. For example, if $M_{BH,min}=2\,M_{\star}=2$ TeV and
$n=6$, the average BH mass is $\langle M_{BH}\rangle\sim 7$ TeV which produces
about seven quanta. If all the quanta are quarks, gluons or gauge bosons, the
number of secondaries produced is $\sim 200$. This number of secondaries is
also close to the mean multiplicity for a SM proton-N$^{14}$ collision with
energy 10$^7$ TeV in the laboratory frame. However, in SM hadronic interactions
most of the momentum is carried by the leading baryon, the other 199 particles
are softer. In the BH case, the momentum is equally shared by all the quanta
such that the shower produces 200 similar secondaries in the first interaction 
which causes a faster shower development.

If the first interaction point could be observed, the difference between the BH
and the SM value of $X_m-X_0$ would be clearly distinguished on an
event-by-event basis (see Fig.\,\ref{longi1} left panel). However, $X_0$ cannot
be directly observed due to  limited sensitivity of the detectors. The right
panel in Fig.\,\ref{longi1} shows the events with the SM curves shifted by 200
g cm$^{-2}$ and renormalized. The difference between the two cases is no longer
apparent.

CC-induced air showers have large fluctuations in $X_m-X_0$ (Fig.\,\ref{longi1}).
This is mainly due to fluctuations in the fraction of primary energy carried by
the scattered electron. This fraction is usually large such that CC-induced
air showers behave often like electromagnetic air showers. On the odd occasion that a
large fraction of the primary energy is carried by partons, the air shower is
closer to a hadronic air shower. In addition, the LPM effect increases the
fluctuations in $X_m-X_0$ of electromagnetic air showers, if the energy of the
scattered electron ($y E_\nu $) is larger than $E_{LPM}$. On the other hand,
the number of particles at shower maximum, $N_{max}$, is proportional to the
primary particle energy, which is more stable in the electromagnetic cascade
case (fluctuations in $N_{max}$ are on the $\sim 5$\% level).

In contrast, BH-induced air showers have small fluctuations in $X_m-X_0$ and
large fluctuations in $N_{max}$. The fluctuations in $X_m-X_0$ are consistent
with the fluctuations of SM hadronic-induced air showers. The large fluctuation
in $N_{max}$ ($\sim 20$\%) is due to the supersposition of two effects: (i)
Each quantum usually carries a large fraction of the primary energy and (ii)
some of the produced quanta do not contribute to the shower energy (neutrinos,
gravitons, $\mu$'s, and non-decaying $\tau$'s).

\begin{figure}
\ybox{0.45}{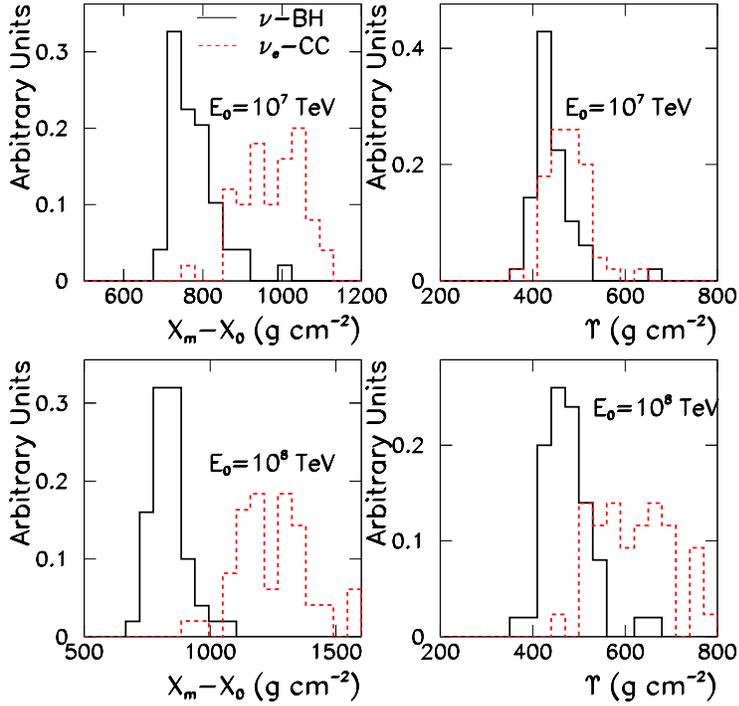}
\caption{The left panels show the $X_m-X_0$ distribution for SM and BH
air showers with $n=6$ and $M_{BH,min}=2\,M_{\star}=2$ TeV. The right panels
show the distribution of the rise-depth parameter, $\Upsilon$, for the same
showers. The upper (lower) panels correspond to $E_\nu=10^7$ ($10^8$) TeV.}
\label{rise}
\end{figure}

\begin{figure}
\ybox{0.35}{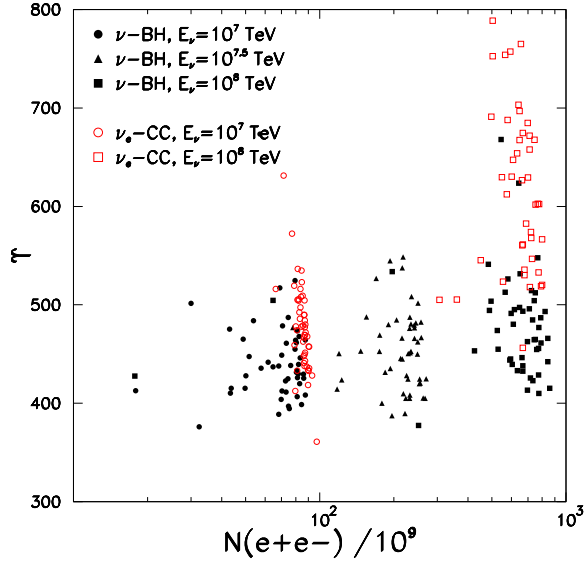}
\caption{Scatter plot of $\Upsilon$ vs $N_{max}$ for CC air showers (red void
symbols) and BH air showers (black filled symbols) with $n=6$ and
$M_{BH,min}=2\,M_{\star}=2$ TeV. The disks are for $E_\nu=10^7$, the circles
for $E_\nu=10^8$ TeV, and the triangles for $E_\nu=10^{7.5}$.}
\label{rise_evol}
\end{figure}

The left panels of Fig.\,\ref{rise} show the $X_m-X_0$ distribution for
neutrino SM and BH air showers. For $E_{\nu}= 10^7$ ($10^8$) TeV, the average
$X_m-X_0$ for BH-induced air showers is 770 (840) g cm$^{-2}$ and 970 (1250) g
cm$^{-2}$  for SM-induced air showers. The spread is 62 (72) g cm$^{-2}$ and 75
(140) g cm$^{-2}$. The difference between the BH and CC air showers increases
with the energy because the difference between hadronic and electromagnetic
air showers also increases with energy (below $E_\nu=10^7$ TeV BH and SM showers
are indistinguishable). Since $X_0$ is not observable, we define an observable
``rise-depth'' parameter for each individual shower,
$\Upsilon \,\equiv\, X_m - X_{0.1}$, where $X_{0.1}$ is the slant depth where
the shower has 10 \% of particle content of the shower maximum. $\Upsilon$ is a
more realistic parameter as a discriminator between BH- and CC-initiated
air showers because it is observable. The right panels of Fig.\,\ref{rise} show
the $\Upsilon$ distributions for SM- and BH-induced air showers at energies
$E_\nu=10^7$ TeV (upper panel) and 10$^8$ TeV (lower panel). The separation
between the distributions is evident at $E_\nu=10^8$ TeV. This trend is better
seen in Fig.\,\ref{rise_evol} where $\Upsilon$ vs $N_{max}$ is plotted for
different primary energies. To clearly see the difference in the distributions
it would be necessary to accumulate a large number of neutrino air showers. For
the cosmogenic neutrino flux, we would expect one $10^8$ TeV neutrino for every
dozen $10^7$ TeV neutrinos \cite{Engel:2001hd}. However, the cosmogenic neutrino
flux is barely detectable by experiments under construction; at most a few events
between $10^6$ and $10^7$ TeV are expected to be detected per year
\cite{Capelle:1998zz,Ave:2000dd,Bertou:2001vm}. Either there are larger
unexpected fluxes of neutrinos or larger detectors will be needed that can
accumulate enough statistics to discriminate between BH and SM interactions
through the $\Upsilon$ distribution.

\begin{figure}
\ybox{0.45}{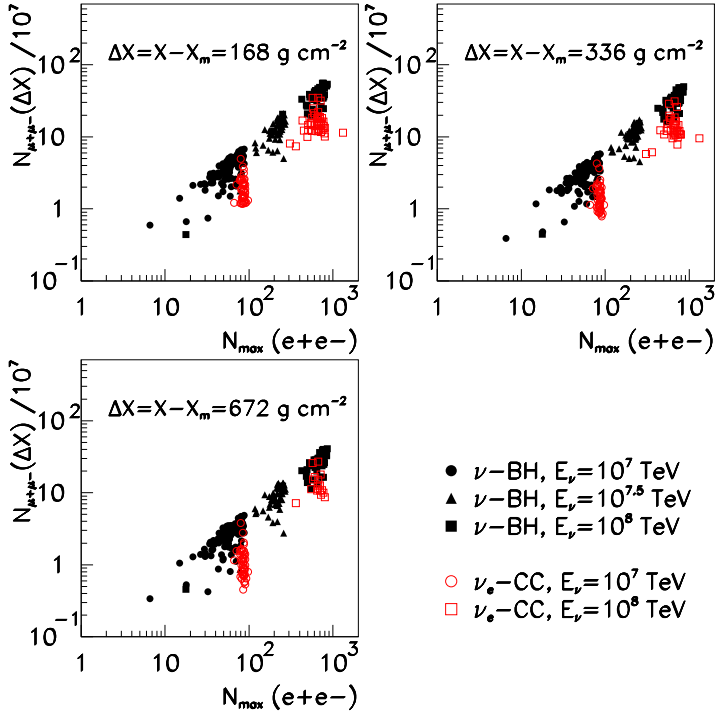}
\caption{$\mu$ number distribution at a depth $X_m+\Delta X$ vs $N_{max}$ for
50 simulated air showers with $E_{\nu} = 10^7$, $10^{7.5}$, $10^8$  TeV. The
red void (black filled) symbols correspond to SM-induced (BH-induced)
air showers ($n=6$ and $M_{BH,min}=2M_{\star}=2$ TeV). $N_{max}$ gives the
observed energy of the event.}
\label{nmu}
\end{figure}

We also simulated the longitudinal development of $\mu$'s for each individual
shower. Since $\mu$'s are detected on the ground, we calculated the $\mu$
number for different positions of the ground detector relative to $X_m$. In
Fig.\,\ref{nmu}, we show the number of $\mu$'s vs $N_{max}$ for 50 air showers
at a depth $X_m+\Delta X$, where $\Delta X=$ 168, 336, and 672 g cm$^{-2}$
$N_{max}$ is essentially proportional to the observed energy. CC-induced
air showers are $\mu$-poor because of their electromagnetic nature, whereas
BH-induced air showers are $\mu$-rich like hadronic air showers.

To summarize, two features should be used to find evidence of BH formation in
extensive air showers: the rise depth and the $\mu$ content of the air showers.
The main differences arise from the electromagnetic nature of the CC-induced
air showers in contrast to the hadronic character of the BH air showers. To
take advantage of the differences in $\mu$ content and the $\Upsilon$
distribution, an experiment should combine both ground and fluorescence
observations for each individual air shower. The Pierre Auger Observatory is
the first such hybrid detector consisting of a ground array which sample the
particle content at a given depth, looked over by four fluorescence detectors
which may determine $X_m$ and $\Upsilon$. For shower core distances larger than
$\sim 1$ km and $\Delta X$ larger than $\sim 100$ g cm$^{-2}$ most of the
signal recorded in the ground detectors is dominated by $\mu$'s and thus is
directly sensitive to the difference in $\mu$ content between BH and CC-induced
air showers. The only challenge for the Auger observatory to test the BH
hypothesis is the low neutrino flux. If the ultrahigh energy neutrino flux is
at the level of the expected cosmogenic flux, a larger version of the Auger
hybrid detector would be needed to test these theories.

\subsection{Shower dependence on BH parameters}
In the previous sections we compared SM- and BH-induced air showers for $n=6$
and $M_{BH,min}=2M_{\star}=2$ TeV. The previous results can be generalized to
different choices of these parameters. The two quantities that characterize the
BH air showers are the cross section and the multiplicity of particles. The
cross section uncertainties considered in \S II affect the first interaction
point, but do not affect the shower development. The main factors that can
change the physical characteristics of the air showers are the multiplicity and the
nature of secondaries originated from the BH evaporation.

The multiplicity is controlled by the mean number of quanta, $N_q$, produced in
the BH evaporation. Most of these quanta are quarks and gluons that hadronize
and initiate a number of hadronic cascades with average energy (laboratory
frame) $E_{q} = \gamma M_{BH}/N_q$. These sub-showers reach a maximum at the
same depth. Thus the maximum of the shower, which is the sum of all
sub-showers, is given by the maximum of a hadronic air shower with energy
$E_{q}$. The shower maximum has a logarithmic dependence on the energy:
\begin{equation}
X_m=X_0+A~{\rm log}_{10}\left[\frac{E_\nu}{N_{q}{\rm TeV}}\right]+B\,.
\end{equation}
The simulations give $A\sim$ 60 g cm$^{-2}$ and $B\sim 311$ g cm$^{-2}$. $A$ is
the change in $X_m-X_0$ per decade of energy and is analogous to the more
commonly used  elongation rate,  which is the change in $X_m$ per decade of
energy. Our results agree well with the experimental results and previous
simulations that give an elongation rate $\sim 60$ g cm$^{-2}$
\cite{Knapp:2002vs}.

The number of quanta depends on the BH mass and differs for each individual
air shower. Moreover, the number of quanta varies with $M_{\star}$ and $n$ at
fixed energy. The shift in the shower maximum of two distinct showers initiated
by $N_{q1}$ and $N_{q2}$ quanta is
\begin{equation}
X_{m1}-X_{m2}=A~{\rm log}_{10} \frac{N_{q2}}{N_{q1}} \ .
\label{nquanta}
\end{equation}
For instance, if the number of quanta increases by a factor of 3 (10), $X_m$
decreases by 29 (60) g cm$^{-2}$.

\begin{figure}
\ybox{0.35}{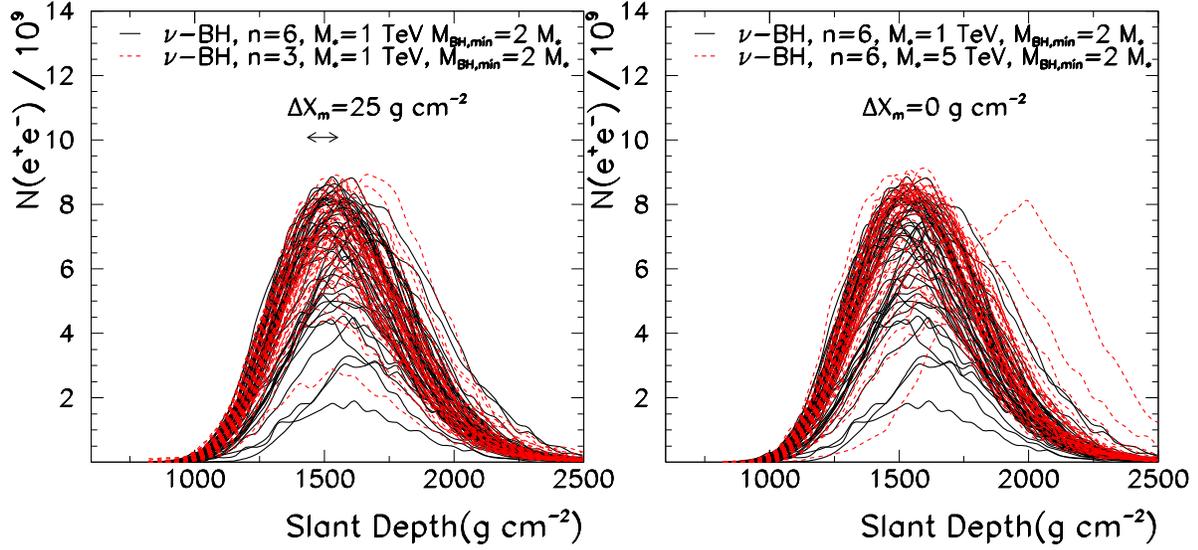}
\caption{Number of $e^+e^-$ vs slant depth for BH-induced air showers. The left
panel shows $n=6$ (black solid lines) and $n=3$ (red dashed lines) for fixed
$M_{BH,min}=2\,M_{\star}=2$ TeV. The right panel shows $M_{\star}=1$ TeV (black
solid lines) and $M_{\star}=5$ TeV (red dashed lines) for
$M_{BH,min}=2\,M_{\star}$ and $n=6$.}
\label{longibh1}
\end{figure}

The left panel of Fig.\,\ref{longibh1} compares the longitudinal development of
BH-induced air showers for  $n=3$ and $n=6$, with $M_{BH,min}=2M_{\star}=2$
TeV. The primary energy is set to $E_\nu=10^7$ TeV, and $X_0$ is the same for
both air showers. The number of quanta produced in the BH evaporation decreases
for larger $n$. Approximately three times more quanta are produced for $n=3$ at
fixed $M_{BH}$. This translates into a shift in $X_m$ of $\sim 25$ g cm$^{-2}$.

The right panel of Fig.\,\ref{longibh1} shows  $X_m$ for $M_{\star}=1$ TeV and
5 TeV with $n=6$ and $M_{BH,min}=2\,M_{\star}$. At fixed $M_{BH}$ the number of
quanta for $M_{\star}=1$ TeV is six times larger than for $M_{\star}=5$ TeV.
However, as $M_{BH}$ is usually slightly larger than $M_{BH,min}$, the 
$M_{\star}=5$ TeV case starts with a more massive BH overall and hence produces
a larger number of quanta than $M_{\star}=1$ TeV  case (see
Fig.\,\ref{probmass}). These two effects counteract and compensate each other,
leading to the same number of quanta for both cases and no shift in $X_m$ as
shown in Fig.\,\ref{longibh1}.

\begin{figure}
\ybox{0.35}{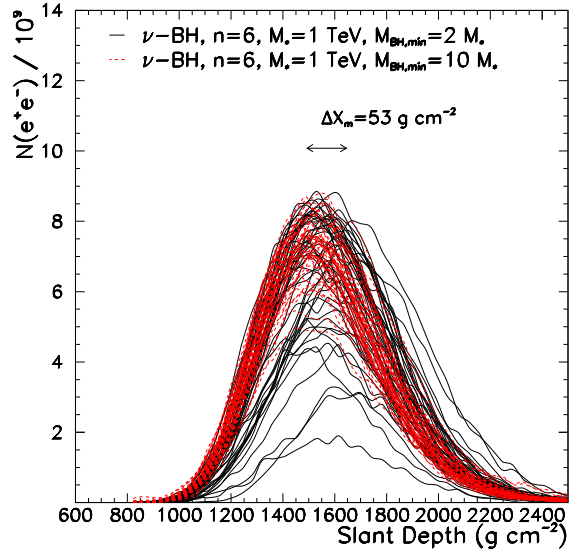}
\caption{Number of $e^+e^-$ vs slant depth for BH-induced air showers with
$M_{BH,min}=2\,M_{\star}$ (black solid lines) and $M_{BH,min}=5\,M_{\star}$
(red dashed lines) at fixed $n=6$ and $M_{\star}=1$ TeV.}
\label{longibh2}
\end{figure}

Fig.\,\ref{longibh2} shows the variation in the longitudinal development for
$M_{BH,min}=2 M_{\star}$ and $10 M_{\star}$. $M_{BH}$ changes by a factor five,
thus increasing the number of quanta by the same factor. This is translated
into a shift in $X_m$ of $\sim 50$ g cm$^{-2}$, in good agreement with
Eq.\,\refb{nquanta}.

The variation of $X_m$ with $M_{\star}$ and the number of dimensions have no
effect on the conclusions obtained in the previous section. The two parameters
discussed to discriminate between BH- and SM-induced air showers do not depend
on the position of the shower maximum. The observable signatures are based on
the difference between the electromagnetic nature of the CC-induced air showers
and the hadronic nature of the BH-induced air showers. Therefore, deeply
penetrating horizontal hadronic-looking air showers will generally signal BH
formation.

\section{Other signatures of BH formation}
In the previous sections we discussed the different characteristics of neutrino
initiated air showers in the atmosphere for SM interactions compared to the
formation of TeV BHs. Given the uncertainties in the BH formation, evaporation
processes, and the inherent fluctuations of air showers, clear signals of BH
formation are difficult to extract and require a large number of neutrino
events. The problem is analogous to separating proton-induced air showers from
gamma-ray air showers, but with an additional unknown $X_0$. If future
experiments can observe both $\Upsilon$ and $\mu$'s of a large number of
neutrino air showers, a separation between SM and BH events could be reached.
Given the low expected flux of ultrahigh energy neutrinos, hybrid
observatories larger than Auger would be necessary.

As an alternative to a large study of neutrino induced air showers, BH
fragmentation may be observable via a few events that have no significant
background. For instance, the production of $\tau$ leptons in BH evaporation
have no significant counterpart in the SM air showers. The fragmentation of
heavy BHs may allow multiple $\tau$ production with $\tau$ energies two or more
orders of magnitude lower than the primary neutrino energy. This kind of
process is strongly suppressed in SM interactions.

One effect of the lower energy of $\tau$'s produced in BH interactions versus
the SM case is the shorter decay length of the generated $\tau$. As a concrete
example, if a neutrino with energy $E_{\nu} = 10^7$ TeV crosses the Andes
mountains towards the Auger Observatory this neutrino-induced BH can produce
one or more $\tau$ leptons with energies around $10^5$ TeV. These $\tau$'s
would decay at a distance of about 5 km from the mountains where the Auger
Observatory is located. A shower from one such $\tau$ decay from the direction
of the Andes would be surprising and even more so if two decays from that
direction were to occur. If the same neutrino had a SM interaction it could
create a single $\tau$ with about $2 \times 10^6$ TeV. This SM produced $\tau$
will decay after traveling about 100 km, past the Auger Observatory.

This example illustrates that for a given neutrino flux and flavor content, the
number of produced $\tau$'s may help separate rare events that have a BH origin
versus a SM origin. Earth-skimming events \cite{Bertou:2001vm, Feng:2001ue}
would also show different energies for the generated $\tau$'s. A significant
study of these signatures depends on detailed assumptions of the neutrino flux
and the detector capabilities and will be more fully addressed elsewhere.

\section{Conclusion}
We considered the possibility of using UHECR observations to detect effects
from TeV gravity theories. We focused our attention on the formation and
fragmentation of BHs at TeV CM energies and found that distinguishing BH
formation and SM air showers is much more challenging than previously expected.

The first challenge on this type of study is the unknown details of BH
formation and fragmentation. The BH formation cross section has large
uncertainties and varies by orders of magnitude with model parameters that
include the number of extra dimensions, the energy scale of extra dimensions,
and the minimum mass of BHs. In principle, contrasting the observed neutrino
flux with the expected neutrino flux can help constrain the neutrino nucleon
cross section, but the uncertainties of the BH cross section limit the
translation of these constraints  into constraints on TeV gravity parameters.

We showed that BH forming interactions generate very different air showers from
SM interactions, but the inability of realistic detectors to observe the first
interaction point hides most of the difference between these air showers. We
proposed two parameters that show the different characteristics of the two
types of air showers: the rise-depth and the muon content of the air showers.
The BH air showers tend to rise faster, given their large multiplicity, and
have larger muon contents, given their hadronic nature. A BH air shower is
similar to a hadronic air shower that can occur at a much higher depth in the
atmosphere, i.e.,  a very deeply penetrating hadronic air shower. Deeply
penetrating SM air showers are dominated by CC processes that generate
electromagnetic air showers. SM neutrino air showers are similar to deeply
penetrating photon-showers. The rise-depth and the muon content can help
distinguish these characteristics of the SM and BH types of air showers, but
the process requires a large number of events to overcome the inherent
fluctuations that generally occur from shower-to-shower. Given that present
observatories are not large enough to study a large number of neutrino events,
the kind of distinction we propose will not be achieved in the near future.

In addition to proposing the study of different air shower characteristics, we
suggested that unique events can arise from BH formation which are suppressed
in SM interactions, such as the multiple $\tau$ generation. The rate for these
events is low if the ultrahigh energy neutrino flux is at the level of the
expected cosmogenic neutrinos. However, unusual air showers from the direction
of a mountain chain can signal both a larger flux of neutrinos and a departure
form the SM interactions.
\begin{acknowledgments}
We are grateful to Jaime Alvarez-Muniz, Luis Anchordoqui, Suyong Choi,
Maria-Teresa Dova, Jonathan Feng, Haim Goldberg, and Al Shapere for useful
discussion and comments. This work was supported in part by the NSF through
grant AST-0071235 and DOE grant DE-FG0291-ER40606 at the University of Chicago
and at the Center for Cosmological Physics by grant NSF PHY-0114422.
\end{acknowledgments}

\end{document}